\documentclass[a4paper,fleqn,usenatbib]{mnras}

\usepackage{newtxtext,newtxmath}
\usepackage[T1]{fontenc}
\usepackage{ae,aecompl}
\usepackage{graphicx}
\usepackage{amsmath}
\usepackage{amssymb}

\title[Stellar Winds and Dense Clumps Around a
Black Hole]{
Relativistic Bondi-Hoyle-Lyttleton accretion in the presence of small rigid bodies around a
black hole}
\author[A. Cruz-Osorio, F. J. S\'anchez-Salcedo and F. D. Lora-Clavijo]{
A. Cruz-Osorio,$^{1,2}$\thanks{aosorio@astro.unam.mx (ACO)},
F. J. S\'anchez-Salcedo $^{2}$\thanks{jsanchez@astro.unam.mx (FJSS)}
and F. D. Lora-Clavijo$^{3}$\thanks{fadulora@uis.edu.co (FDLC)}
\\
$^{1}$Frankfurt Institute for Advanced Studies, Ruth-Moufang-Stra{\ss}e 1, D-60438 Frankfurt am Main, Germany\\   
$^{2}$Instituto de Astronom\'{\i}a, Universidad Nacional Aut\'{o}noma de M\'{e}xico, AP 70-264, 
Mexico City 04510, Mexico \\ 
$^{3}$Grupo de Investigaci\'on en Relatividad y Gravitaci\'on, Escuela de F\'isica, Universidad Industrial de Santander, A. A. 678,\\
Bucaramanga 680002, Colombia\\
}

\date{Accepted XXX. Received YYY; in original form ZZZ}

\pubyear{2015}

\begin{document}
\label{firstpage}
\pagerange{\pageref{firstpage}--\pageref{lastpage}}
\maketitle

%% ABSTRACT %%
\begin{abstract}

We study the relativistic Bondi-Hoyle-Lyttleton accretion onto a Schwarzschild black hole (BH), which is
surrounded by rigid and small, randomly distributed, bodies. These bodies are idealized 
representations of substructure like stars passing close to the BH, bubbles created by stellar winds 
or cold clumps. We explore cases where the filling factor of these bodies is small.
The flow is assumed to be adiabatic and move supersonically towards the black hole.
The interaction with these rigid obstacles transforms ram pressure of the
flow into thermal pressure through bow shocks, slowing down the flow
and making the accreting gas turbulent. As a consequence, although
the flow reaches a statistically-steady state, the accretion rate presents 
some variability. For a flow Mach number at infinity of $4$, a few of these
objects ($5-10$) are enough to increase the accretion rate about $50 \%$ over
the accretion rate without bodies, even though the gas is adiabatic 
and the filling factor of the obstacles is small.

\end{abstract}

%% KEYWORDS %%

%% INTRIDUCTION %%
\section{Introduction}
Gas accretion on to a gravitational mass, e.g., stars, compact objects or black holes (BHs),
is of great interest in many areas of astrophysics.
In particular, gas accretion on to BHs is associated with the formation and growth of 
supermassive BHs (SMBHs), X-ray transients and gamma-ray bursts (GRBs).  
Many investigations have been undertaken to understand the different aspects of BH accretion \citep[see][for recent reviews]{2005Sci...307...77N,2013LRR....16....1A, 2014ARAA..52..529Y}.

The Event Horizon Telescope (EHT)\footnote{http://www.eventhorizontelescope.org}
will be able to resolve the BH structures on event horizon scales,
which may be crucial not only to test General Relativity \citep{2014ApJ...784....7B}
but also to study accretion processes very close 
to the BH event horizon \citep{2015MNRAS.446.1973R}. In fact, the SMBH hosted in the centre of our Galaxy, 
with a mass of $M = (4.1\pm 0.6) \times 10^{6}M_{\odot}$ \citep{2008ApJ...689.1044G}, 
can be used as a laboratory to understand the interaction 
between it and the gaseous and stellar environment. For instance, \cite{2008Natur.455...78D}  have reported observations at a wavelength of 1.3 mm that set a size of $37^{+16}_{-10}$
 microarcseconds on the intrinsic diameter of Sgr A*.  This value suggests that the bulk 
emission of Sgr A* arises from the surrounding accretion flow.
Moreover, \citet{2011ApJ...727L..36F} found variability in the flux density in
$1.3$ mm VLBI observations of Sgr A*, which was interpreted as a brightening of structure on 
scales of a few Schwarzschild radii.
Recent observations have shown evidence for partially ordered fields near the 
event horizon, on scales of $~6$ Schwarzschild radii \citep{2015Sci...350.1242J}.   
Based on these observations,  it is crucial to understand how the accretion on to BHs occurs
at scales of the event horizon. 

Accretion on to a star moving supersonically through a homogeneous medium was studied by 
Bondi, Hoyle and Lyttleton \citep[][hereafter BHL]{1939PCPS...35..405H,1944MNRAS.104..273B}. Later, \cite{1952MNRAS.112..195B} studied  the spherical accretion of gas on to a central gravitational object. 
In the BHL formalism, the mass accretion rate is given by
\begin{equation}
\dot{M}_{\rm BH}=\frac{4\pi G^{2}M^{2}\rho_{\infty}}{(v^{2}+c_{s}^{2})^{3/2}},
\label{eq:mdotexact}
\end{equation}
where $M$ is the mass of the BH which moves at velocity $v$ relative
to the otherwise homogeneous ambient medium, which has density $\rho_{\infty}$ and sound 
speed $c_{s}$.  
This formula has applications in many areas, for example, to estimate the accretion
rate by a compact star in a binary system \citep[e.g.][]{1978ApJ...224..625P}, or to estimate the gas 
supply to central supermassive BHs (SMBHs) in cosmological simulations 
\citep{2005MNRAS.361..776S, 2012ApJ...754...34J, 2013MNRAS.429.2594B}.

There are many ingredients that can change the mass accretion rate on to the BH with
respect to the canonical value given in Equation (\ref{eq:mdotexact}): the physics of stellar feedback, 
AGN feedback, the presence of small-scale structure, turbulence and magnetic fields in the multiphase
interstellar medium, accretion-disc winds, the accretion of stars, and so on \citep[e.g.,][]{2016MNRAS.458..816H}.

The role of turbulence has been studied by
\citet{2006ApJ...638..369K}, who derive an estimate of the distribution of accretion rates on to
point particles in supersonically turbulent gas. They find that accretion 
is bimodal; the flow pattern around some of the particles resembles the classical Bondi-Hoyle flow,
but the flow pattern around other sink particles is closer to the case of high vorticity.
In the latter case, vorticity inhibites accretion \citep{2005ApJ...618..757K}. Nevertheless,
the mean accretion rate can be significantly enhanced with respect to the value predicted by the 
unmodified Bondi-Hoyle formula.

The standard Bondi-Hoyle prescription for
the accretion rate also ignores the rotation of the gas, which may lead to the formation
of an accretion disc around the BH. Strictly, it is necessary to resolve the accretion radius,
defined as $r_{a}=GM/(v^{2}+c_{s}^{2})$, to have
an accurate estimate of the accretion rate. Some attempts have been made to give an 
accretion rate prescription taking into account the gas angular momentum \citep[e.g.][]{2011MNRAS.412..269P}. 
\cite{2011MNRAS.415.1027H} developed a sub-grid model, which can be incorporated in cosmological simulations, based on analytical calculations
about angular momentum transport and gas inflow. A novel refinement scheme has been
presented in \citet{2015MNRAS.454.3445C} which improves the estimation of the gas properties
close to the accretion radius \citep[see also][]{2016MNRAS.463...63C}.  \citet{2015MNRAS.454.1038R} explore the effect of accounting for the angular momentum
by using an improved accretion model that includes the circularization and 
subsequent viscous transport of infalling material.
On the other hand, a sub-grid model that
distinguishes between hot and cold gas accretion has been implemented by \citet{2015MNRAS.448.1504S}.

The role of stellar feedback and AGN feedback has been investigated in several
papers \citep{2006MNRAS.366..358C, 2008MNRAS.383..458C, 2013MNRAS.430.1970H,
2016MNRAS.456L..20B,2017MNRAS.467.3475N}.
More specifically, \citet{2006MNRAS.366..358C, 2008MNRAS.383..458C} simulate the stellar wind dynamics
within the central parsec of the Galactic centre and find that a variable accretion rate is produced by the 
infall of cold clumps formed by colliding slow winds. Still, they argue that the total amount of
accretion is dominated by hot gas. 
On the other hand, \citet{2013MNRAS.430.1970H} suggest that in cases such as in the
galaxy NGC 3115, the stellar winds of stars orbiting the BH may shock and produce an over-pressured 
hot bubble that suppresses the inflow of ISM gas. These authors also
suggested that under certain conditions, the gas can radiatively cool and form cold clumps 
that may be accreted by the BH. More recently, the numerical simulations by \citet{2016MNRAS.456L..20B}
show that feedback from multiple nuclear stellar sources may lead to
the formation of high density clumps and filamentary structures that are difficult to
expel. They suggest that this cold material could fall back to the nuclear cluster and
form new stars or feed the central BH.
Using simulations with a resolution of $0.1$ pc, \citet{2017MNRAS.467.3475N} find that
AGN feedback can keep the BH in a self-regulation regime, and show that Bondi subgrid
algorithms in low resolution simulations can both under- and over-predict the actual
accretion rates. 
Other numerical experiments indicate that dense cold filaments of gas are still
able to flow into the inner parts of a galaxy even in the
presence of AGN feedback \citep{2014MNRAS.441.3055B} or supernova feedback 
\citep{2015MNRAS.452.3593H}.
Dense filaments and clumps are not only formed by converging stellar winds or supernova bubbles;
thermal instabilities may develop in the accretion flow, forming dense clumps and leading to
subsequent chaotic accretion \citep{2011MNRAS.418..591B,2013MNRAS.432.3401G}. 
Recent observations lend support to this clumpy accretion \citep{2016Natur.534..218T}.

To reach present day masses of massive BHs in the centres of galaxies,
BHs must have been able to accrete material (gas and stars) in a very efficient way \citep{2010A&ARv..18..279V}. Moreover, \cite{2011Natur.474..616M}  reported the observations of a quasar  
at a redshift of $7$, which hosts a SMBH with a mass of $2\times10^9M_{\odot}$, implying  
that the masses of high redshift SMBHs make the requirement of efficient accretion processes even greater at high redshift. Since central BHs are generally embedded in a nuclear star cluster \citep[][and references therein]{2015ApJ...803...81N}, tidal captures of stars may provide an additional channel for BH mass growth
\citep[e.g.][]{2017MNRAS.467.4180S}.

Recent surveys have shown that the luminosity at large energies in close
dual active galactic nuclei (AGNs) is enhanced as compared to their isolated counterparts (e.g., Liu
et al. 2013 and references therein). \citet{2015ApJ...803...81N} have explored why accretion rates may
be increased during galaxy mergers despite the fact that BHs travel supersonically through the surrounding gas.
They show that gravitational focusing effects of nuclear star clusters can lead to an
enhancement in SMBH accretion rates up to an order of magnitude, even though this system (SMBH plus 
nuclear star cluster) moves with a Mach number of $\sim 1.3-1.7$ relative to the gas.

In the present work, we show, for the first time, a model of the relativistic BHL accretion 
of hot gas, in the presence of small obstacles around the accretor. 
In the astrophysical context, relativistic BHL accretion  has been used to study different
processes, from the vibrations in the shock cone, which may potentially
explain high energy phenomena like QPOs \citep{2011MNRAS.412.1659D, 2012MNRAS.426.1533D, 2013MNRAS.429.3144L}, to ultra-relativistic processes associated with the growth of primordial 
BHs during the radiation era \citep{2013MNRAS.428.2171P,2013AIPC.1548..323C}.
Given the importance of understanding the accretion process, 
the relativistic BHL problem has also been studied in the presence of magnetic 
fields \citep{2011MNRAS.414.1467P}, radiative terms \citep{2011MNRAS.417.2899Z}, and with density or velocity gradients \citep{2015ApJS..219...30L, 2016arXiv160504176C}.

Motivated by the fact that one cannot assume that BHs at the galactic centres are naked
accretors, we have investigated, in the relativistic regime, how the accretion rate and structure of the wake around 
a BH are modified by the presence of small and rigid `obstacles'.
These obstacles are idealizations of small-scale substructure that may be present in the flow,
such as dense and cold clumps, passing stars, stellar debris or stellar bubbles.
When a gas streamline encounters a rigid substructure, 
the gas shocks and the kinetic pressure thermalizes, 
changing completely the relativistic BHL picture of accretion where gas can only shock
along the Mach cone behind the accretor (the so-called accretion column). 

The paper is organized as follows. In Section \ref{sec:ENS}, we present equations, the numerical methods and
the details of our initial conditions. In Section \ref{sec:SFAR}, we describe the structure of the flow and accretion rates from our numerical simulations. Finally, we  summarize our main results in Section \ref{sec:conclusions}. 
It is worth mentioning that, unless otherwise stated, we use natural units
in which $G=c=1$ and thus length and time are measured in units of $M$.

%% EQUATIONS AND NUMERICAL SETUP%%
\section{Relativistic Equations, numerical methods and initial conditions }
\label{sec:ENS}

\subsection{Relativistic hydrodynamic equations}

In order to study the relativistic Bondi-Hoyle accretion onto a BH, we solve the equations that describe the evolution of a relativistic perfect fluid in a curved space-time, which can be written down from the conservation of the rest mass density $\rho$ and the local conservation of the energy momentum tensor $T^{a b}$
\begin{eqnarray}
\nabla_{a} (\rho u^{a})=0, ~~~~~~~ \nabla_{b}T^{a b}=0,
\end{eqnarray}
where $u^{a}$ are the components of the 4-velocity of the fluid elements and $\nabla_a$ stands for the covariant derivative associated with the 4-metric $g^{ab}$. To model the relativistic gas, we use the ideal equation 
of state, which is given by 
\begin{eqnarray}
e = \frac{p}{(\Gamma - 1)\rho},
\end{eqnarray}
where $e$ is the specific internal energy density, $p$ the pressure of the fluid and $\Gamma$ the adiabatic index
\citep[for more details, see][]{2013MNRAS.429.3144L,2015ApJS..219...30L}. 
These equations are solved onto a non-rotating BH, which is described by Schwarzschild line elements in horizon penetrating Eddington-Finkelstein coordinates as
\begin{eqnarray}
ds^2 = -\alpha^2dt^2 + \gamma_{ij}(dx^i + \beta^i dt)(dx^j + \beta^j dt) , 
\end{eqnarray}
where the lapse function $\alpha$, the shift vector $\beta^i=(\beta^r,0,0)$ and the spatial metric $\gamma_{ij}$ are
\begin{eqnarray} 
\nonumber \alpha&=&\left(1+\frac{2M}{r}\right)^{-1/2}, ~~~~~~
\beta^i=\left[ \frac{2M}{r}\left(1+\frac{2M}{r}\right)^{-1},0,0\right], \\
\gamma_{ij}&=&{\rm diag}\left[1+\frac{2M}{r},r^2,r^2\sin^2\theta \right].
\end{eqnarray}
In this work, we consider the test field approximation, that is, we do not evolve the geometry of the space-time, but
keep our background space-time fixed.

\subsection{Numerical methods and initial conditions}
We solve the relativistic hydrodynamic equations onto a curved spacetime of a Schwarzschild BH, 
using CAFE,  a fully three-dimensional relativistic-magnetohydrodynamic code 
\citep{2015ApJS..218...24L}. Although the MHD in CAFE is written for a
Minkowski space-time, the hydrodynamical (HD) routine may employ either equatorial \citep{2012MNRAS.426..732C} or axial symmetry \citep{2013MNRAS.429.3144L},  in order to 
include fixed curved space-times. Specifically, 
to integrate the hydrodynamical equations, we employ the HRSC method with
HLLE flux formula in combination with the minmod linear piecewise variable reconstructor. 

Our simulations are performed in a spherical grid $(r,\theta)$, centered in the BH, assuming axial symmetry
\citep[for more details about the code, convergence and other
tests, see][]{2013MNRAS.429.3144L}. The velocity
components of the initial flow are chosen to represent a homogeneous gas which
fills the whole domain and moves along the $z$-direction with velocity $v_{\infty}$.
Therefore, the initial velocity in Cartesian coordinates is $v^{x}=v^{y}=0$ and
$v^{z}=v_{\infty}$. However, in our coordinate system, we have 
$v^{r}=v_{\infty}\cos(\theta)/\sqrt{\gamma^{rr}}$ and $v^{\theta}=-v_{\infty}\sin(\theta)/\sqrt{\gamma^{\theta \theta}}$, where  $\gamma^{\theta \theta}$ and $\gamma^{\theta \theta}$ are the inverse metric components \citep[for more details, see][]{2013MNRAS.429.3144L, 2012MNRAS.426..732C}.
In order to specify the initial velocity of the flow, we use the asymptotic values of
the Newtonian Mach number ${\mathcal{M}}_{\infty}$ and the sound speed $c_{s,\infty}$ (note that
in the relativistic case, the Mach number alone does not specify the flow).
We assume a gas adiabatic index $\Gamma=5/3$, appropriate for accretion without radiative losses,
a specific value for the sound speed $c_{s,\infty}=0.1$, and explore models with ${\cal M}_{\infty}$ 
between $4$ and $5$.

In our spherical grid, the computational domain is
$\theta \in [0,\pi]$ and $r\in [r_{\rm exc},r_{\rm max}]$, 
where $r_{\rm esc}$ and $r_{\rm max}$ define the inner and outer boundaries, respectively.
The inner boundary or excision boundary $r_{\rm exc}$ is located inside the event horizon of the BH
\citep[for more details, see][]{2012MNRAS.426..732C}. This is a natural choice since no physical
information may propagate from inside the event horizon to outside.
Gas going inside $r_{\rm exc}$ is removed from the simulation.
The accretion physics for a naked BH is characterized by the accretion radius (we remind that
$r_a = M / [c_{s,\infty}^2 ( 1+{\cal M}_{\infty}^2 )]$ in our geometrized units). 
This implies that we need to have enough numerical
resolution to solve the accretion radius. Moreover, in order to simulate the interaction of
a BH moving in a homogeneous medium, it is usually required that 
the distance of the BH to the upstream boundary must be larger than
$\sim 4r_{a}$ \citep[e.g.][]{1994ApJ...427..351R}. This condition imposes a minimum size for the
computational box. For the outer boundary, we take $r_{\rm max}= 10.5 r_{a}$. It is worth mentioning that, in our geometrized units, 
the accretion radius for the Newtonian Mach numbers $4$ and $5$ are $5.88M$ and $3.85M$, respectively.
In our code units, we take $M=1$, implying accretion radius of $5.88$ and $3.85$, respectively. 
In all the models we use the same resolution: 
$\Delta r = 7.81\times 10^{-2}$ and $\Delta \theta = 1.22\times 10^{-2}$.

In order to compute the mass accretion rate onto the BH, we use a spherical detector 
at the event horizon of the BH, evaluating the expression
\begin{equation}
\Dot{M} = -2\pi \int \rho r^2  v^r  \sin \theta d \theta.
\label{eq:mrate}
\end{equation}
Although the BH should increase its mass due to mass
accretion, we do not add the gas mass to the BH because the change in the BH mass is 
usually very small.

We include the presence of $N_{b}$ rigid and static bodies in our domain in order to study the accretion flow under the presence of obstacles.
Along $R-z$ cross sections, these rigid obstacles are assumed to be circular, with fixed radius $r_{b}$. For conciseness, we will refer to them indistinctly as ``clouds'', ``clumps" or ``rigid bodies'',  although the imposed axial symmetry implies that they have torus shape in the 3D space. We use reflecting boundary conditions on the surface of the clouds.
We are interested in a regime where the clouds are small as compared to the accretion
radius of the BH. We assume $r_{b}=0.5M$, which implies that 
the radius of the clouds are smaller by a factor of $\geq 8$
than the BH accretion radius, for the values of ${\cal M}_{\infty}$ 
under consideration.

We ignore the gravitational influence of the clouds, i.e., the metric of the BH
is fixed along the evolution and is given by the Schwarzschild solution. 
Neglecting the gravity of the clouds is justified if two conditions are satisfied:
(1) the total mass in the clouds $N_{b}M_{b}$ is $\ll M$ and (2) the accretion radius 
of each individual cloud $\simeq M_{b}/(c_{s,\infty}^{2}{\mathcal{M}}_{\infty}^{2})$ is $\ll r_{b}$.
Combining with our assumption that $r_{b}=0.5M$, the latter condition implies that $M_{b}\ll 0.08M$
for ${\cal M}_{\infty}=4$.

We take all the clouds to be identical and randomly distributed in our numerical domain. 
It is expected that the structure of the flow will depend sensitively on $r_{b}$ and on the filling factor of the clouds, 
defined as $\eta\equiv N_{b}(r_{b}/r_{\rm max})^{2}$ (remind that $r_{\rm max}$ is the external radius
of our domain). For instance,
in the hypothetical case that the filling factor is close to $1$, the gas can only flow through 
tiny cavities between clouds. We are not interested in this case but 
in a regime where the filling factor is small. 
We explore three different values for $N_{b}$: $0, 5$ and $10$. Consequently, 
$\eta\leq 2\times 10^{-3}$. Although the position of the clouds were drawn randomly, all the
simulations with $N_{b}=5$ have the clouds at the same positions to facilitate comparison.
In the simulations with $N_{b}=10$, we put $5$ clouds in the same positions and add the
remaining $5$ clouds at arbitrary positions.

%%%%%%%%%%%%%%%%%%%%%%%%%%%%%%%%%%%%%%%%%%%%%%%%%%%%%%%%%%%%%%%%%%%%%%%
\section{Structure of the flow and accretion rates}
\label{sec:SFAR}
%%%%%%%%%%%%%%%%%%%%%%%%%%%%%%%%%%%%%%%%%%%%%%%%%%%%%%%%%%%%%%%%%%%%%%%
 
\begin{figure*}
\includegraphics[width=0.57\columnwidth]{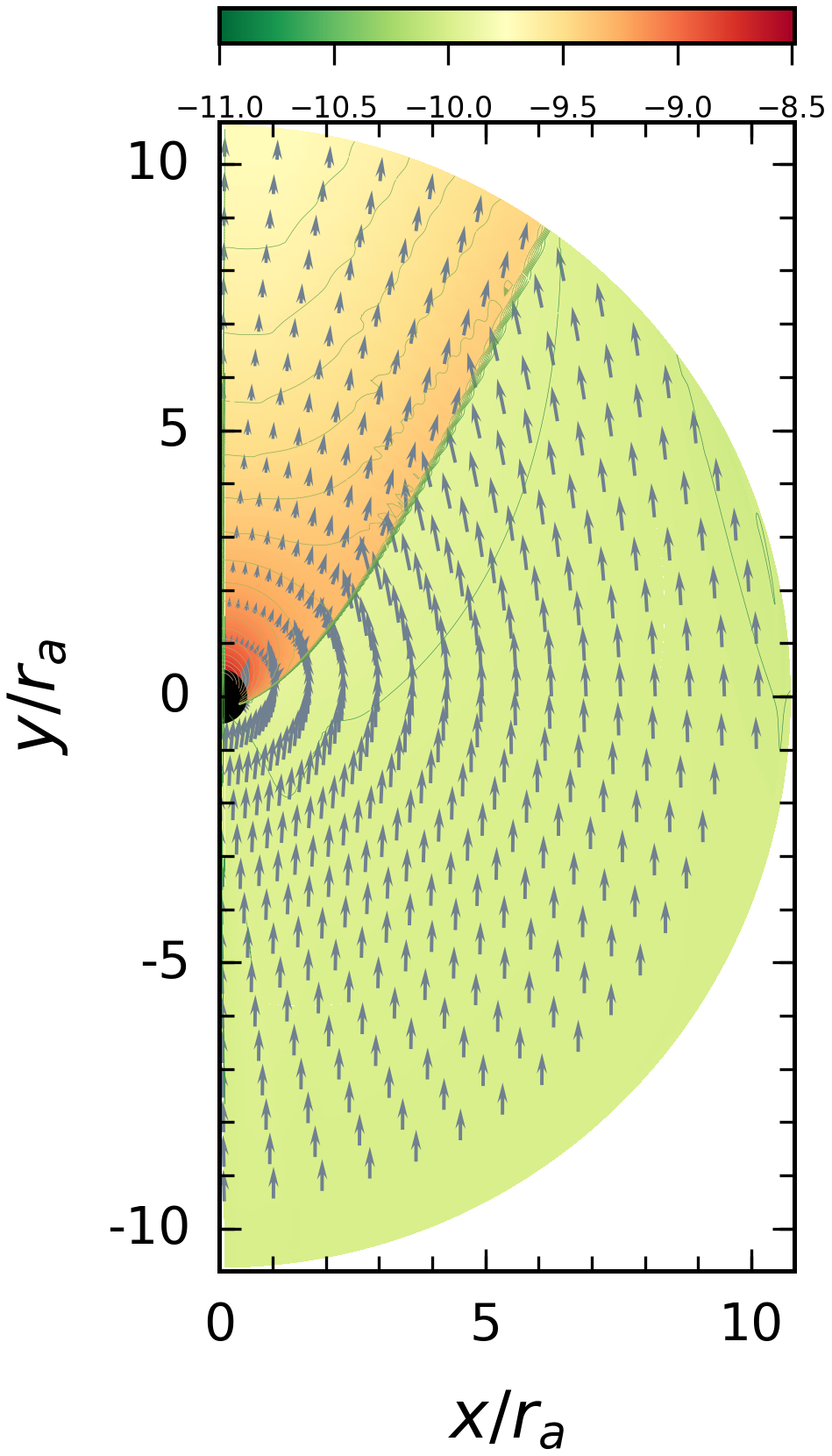}\includegraphics[width=0.5\columnwidth]{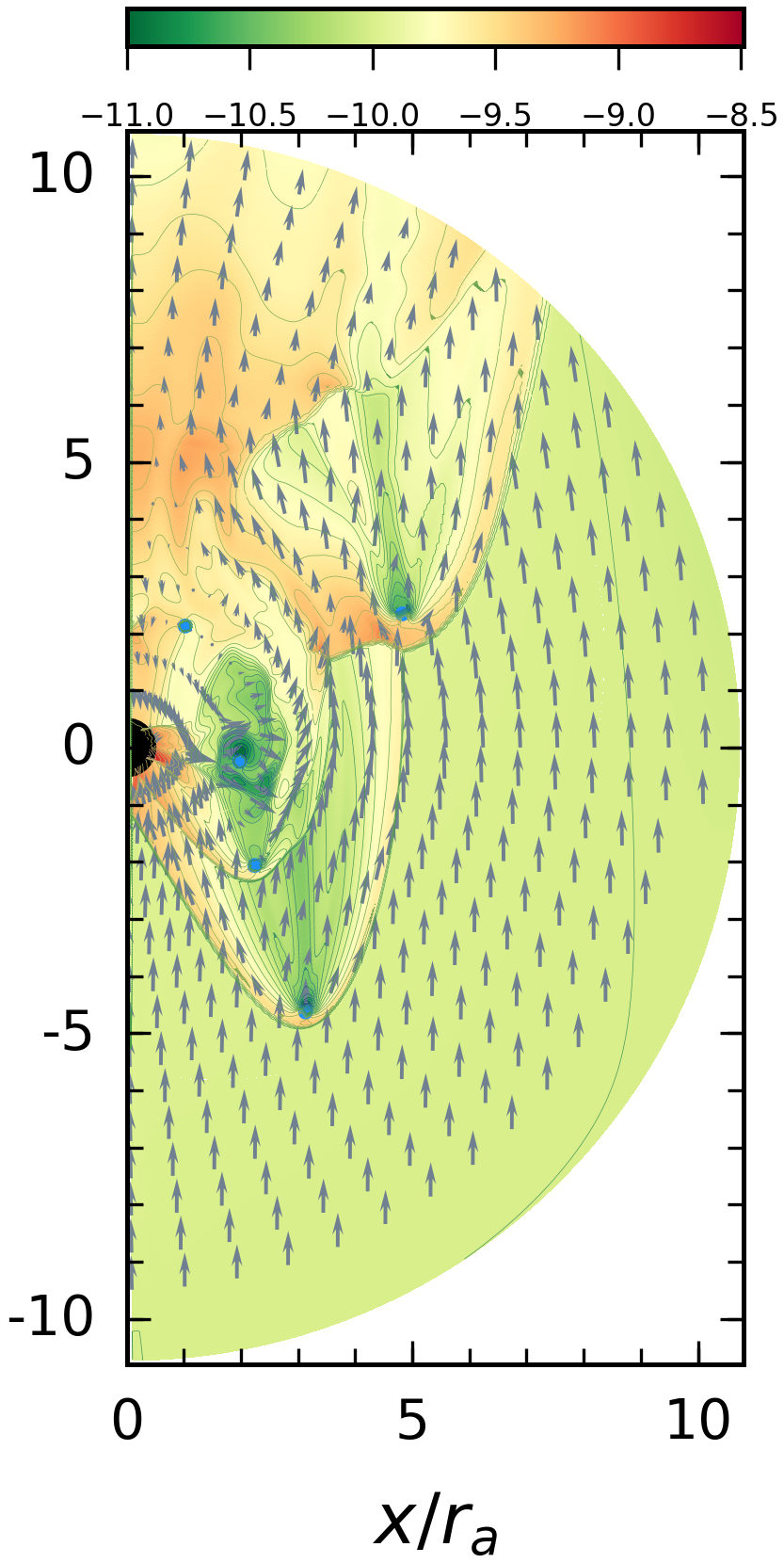} \includegraphics[width=0.5\columnwidth]{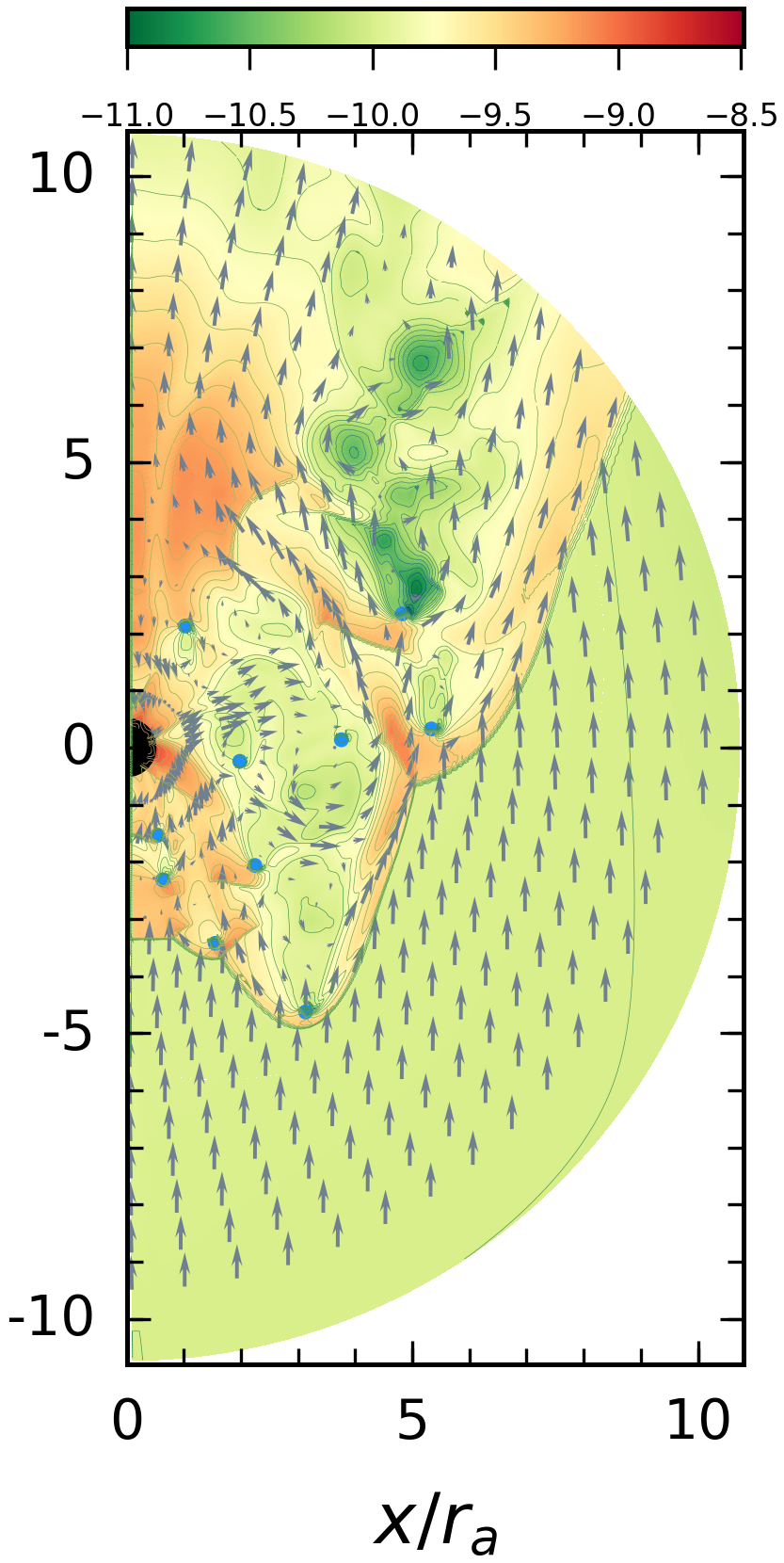}\includegraphics[width=0.5\columnwidth]{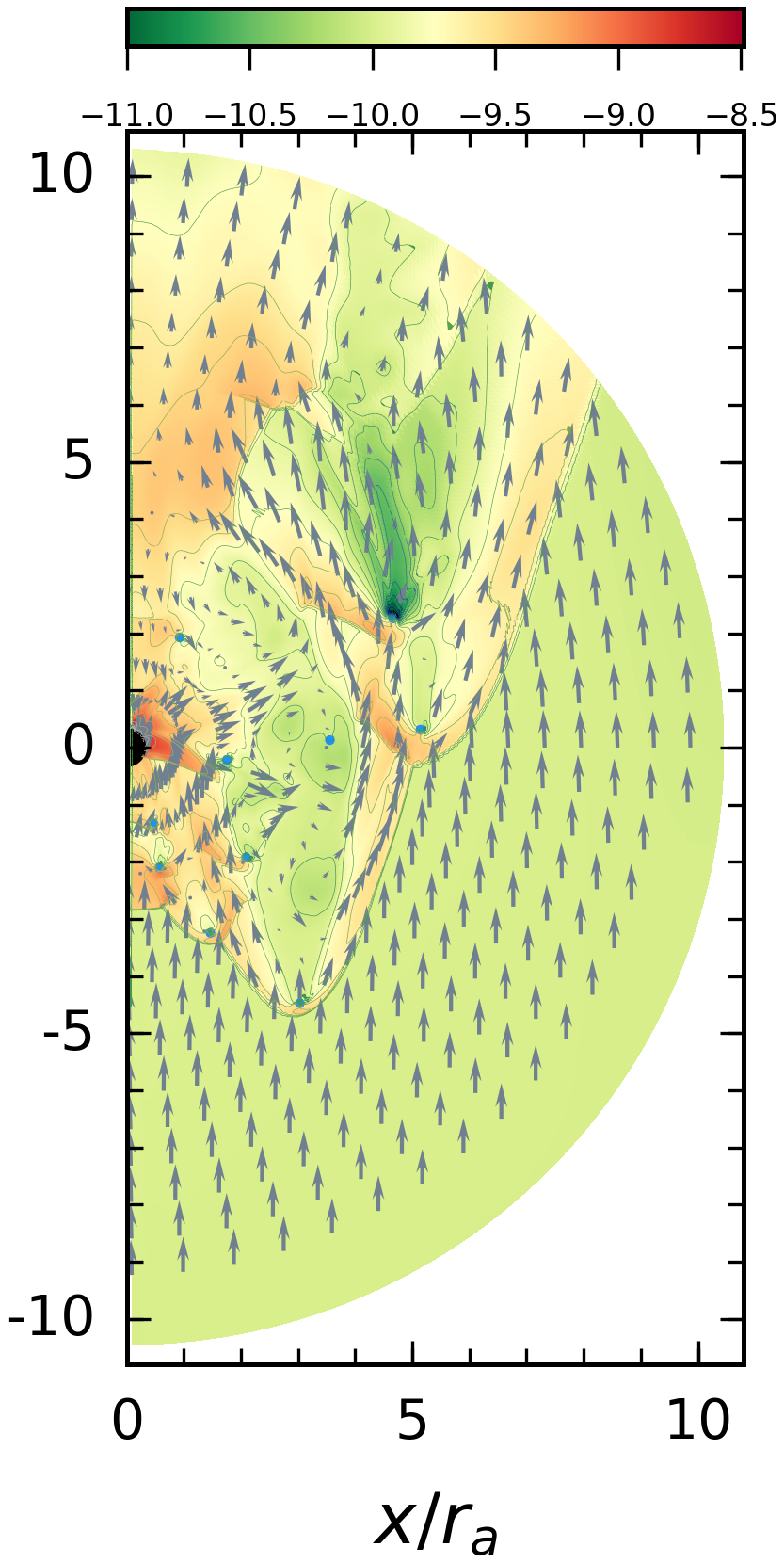}\\
\includegraphics[width=0.57\columnwidth]{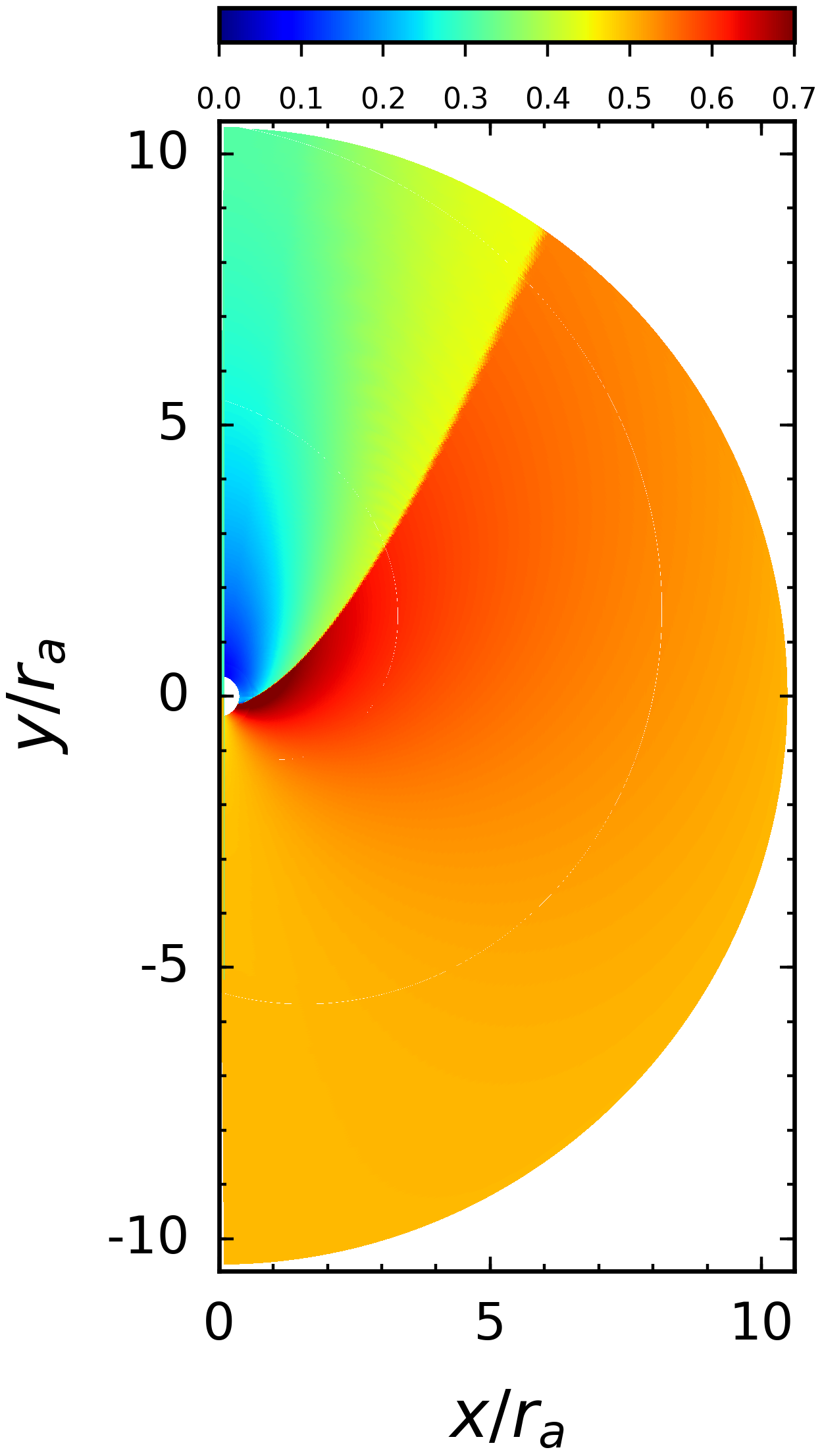}\includegraphics[width=0.5\columnwidth]{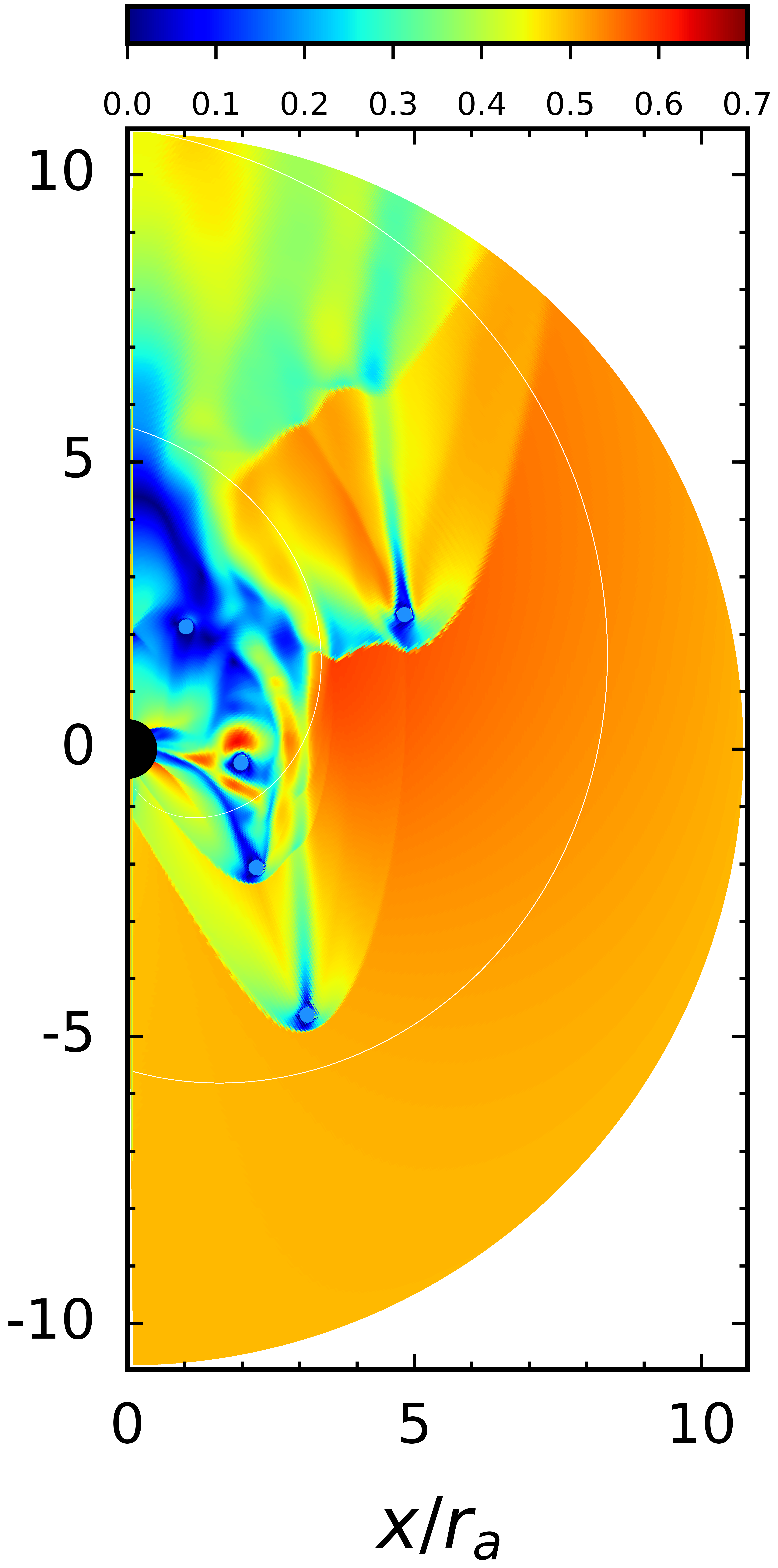} \includegraphics[width=0.5\columnwidth]{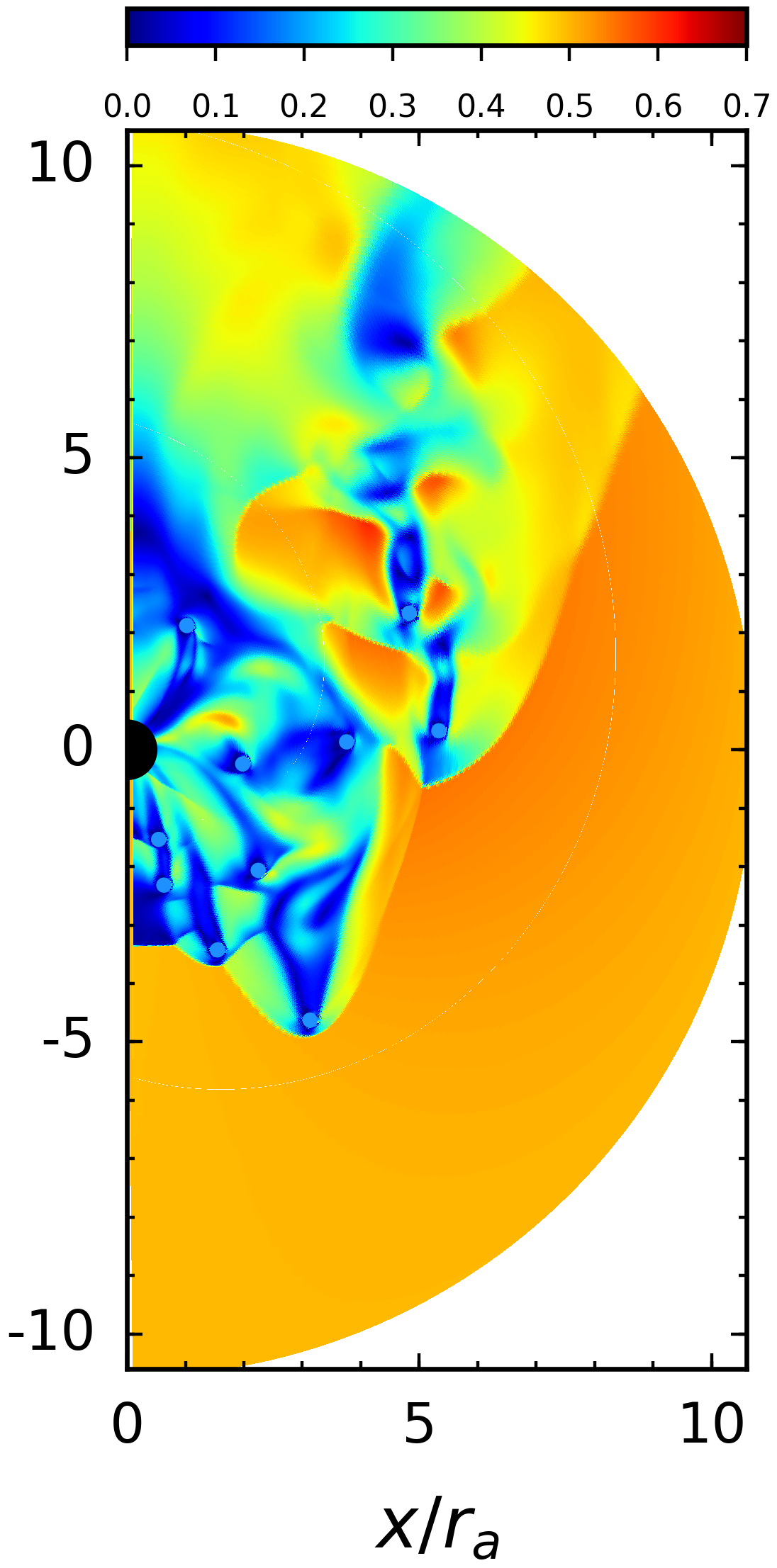}\includegraphics[width=0.5\columnwidth]{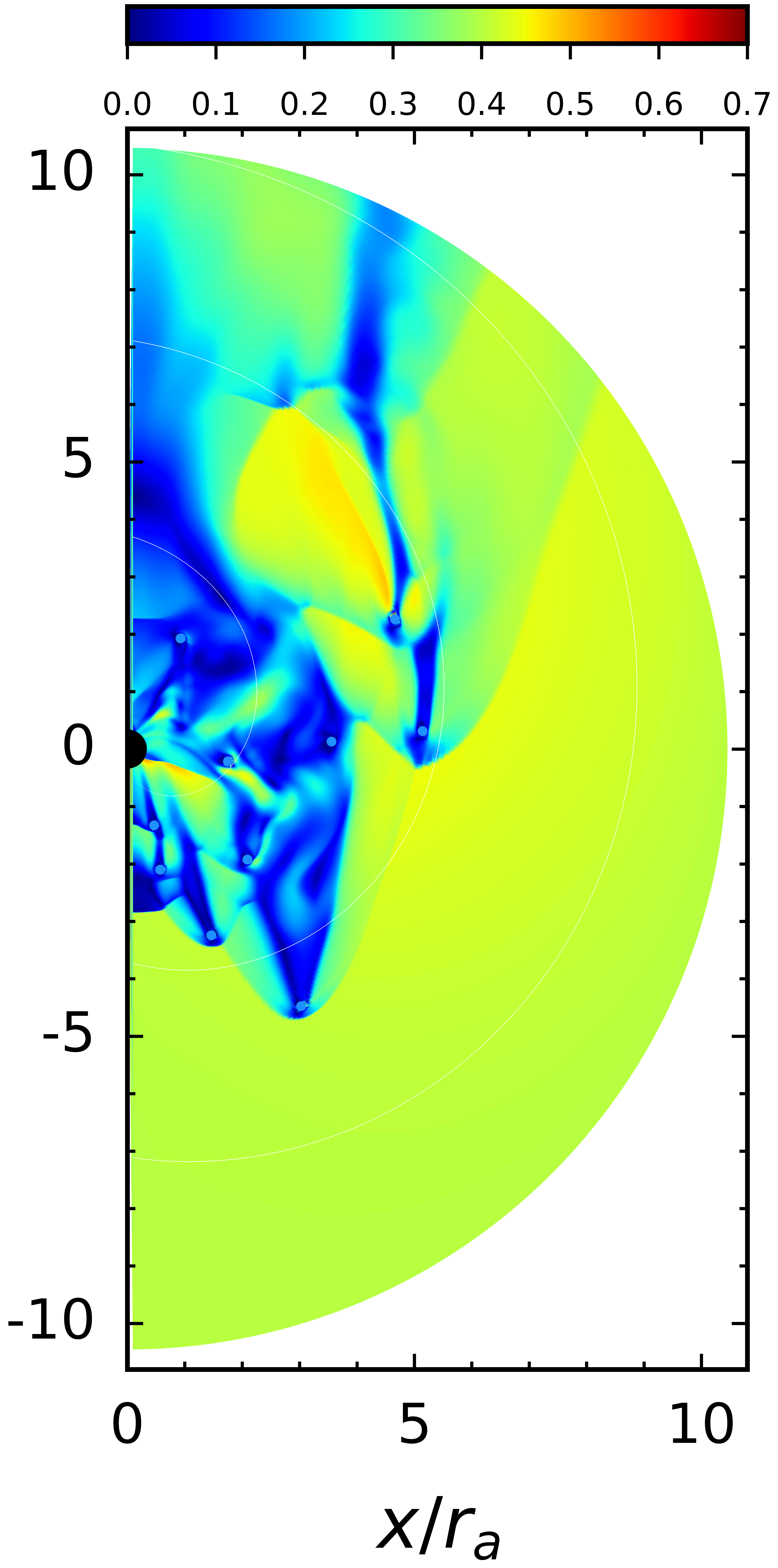}\\
\caption{Morphology of the rest mass density in logarithmic scale (top) and the magnitude of the velocity (bottom) around a Schwarzschild BH at $t=1000M$, for ${\cal M}_{\infty}=5$ and $N_{b}=0$ (left row), for ${\cal M}_{\infty}=5$ and $N_{b}=5$ (second row) for ${\cal M}_{\infty}=5$ and $N_{b}=10$ (third row) and for ${\cal M}_{\infty}=4$ and $N_{b}=10$ (right row).  The black point is the black hole and the blue points are the positions of the clumps. In these plots, the axes are in units of $r_{a}$. Clearly visible are the low-density regions behind the clouds and the formation of bow shocks.  In all cases, the Mach cone at the rear of the accretor is not as well defined as in the classical 
BHL flow.  For ${\cal M}_{\infty}=5$ and $N_{b}=10$, the structure of the gas is more complex.}
\label{fig:bnumber}
\vspace{0.85cm}
\end{figure*}

At early times in the simulation, since the flow is supersonic, each cloud creates an individual bow shock front where the gas density increases and the flow decelerates. Low-density regions past the clouds are also visible (see the first row of Fig. \ref{fig:bnumber}) where the fluid is moving with very low velocities (see the second row of the Fig.  \ref{fig:bnumber}). At later times, the bow shocks of different clouds merge and a large-scale bow shock can be seen in some cases. The interaction between individual bow shocks produces new shock fronts and 
fluctuating structures in the density and also in the
velocity field of the flow, but a statistically steady flow is achieved after $\sim 200 M$. 
In the presence of obstacles, the flow becomes more chaotic than it is without clouds;
swirling motions can be observed in Figure \ref{fig:bnumber}.
As expected, the flow becomes more chaotic for larger $N_{b}$ (see Fig. \ref{fig:bnumber}).  As a consequence of the perturbations created by the clouds, the classical Mach cone at the rear of the accretor becomes difficult to trace. Indeed, if the number of clouds is large enough, the Mach cone is completely erased and, instead, a subsonic envelop around the accretor could be formed.

Since the pioneering work of \cite{1939PCPS...35..405H}, it is common to formulate
the problem in terms of the accretion radius defined as the maximum impact parameter
for which streamlines will be accreted. In the absence of clouds, the accretion
radius may be estimated using the ballistic approximation, which ignores fluid
effects. In the presence of clouds, one expects that the accretion radius for 
supersonic flows will become larger as soon as the filling factor is small. 
The reason is the following. When a streamline with $v_{z}>0$ collides with a rigid body,
we have that $dv_{z}/dt<0$ and this reduction of the relative velocity of the
flow with respect to the BH may lead the material becoming bound to the BH. 
As a result, the effective area for accretion 
increases when the obstacles are added. In addition to this, the
obstacles can redirect the flow towards the BH. These two effects promote
accretion onto the BH. Those obstacles that are expected to contribute
most to enhance accretion on to the BH are those situated at distances from the BH 
larger but comparable to the accretion radius $r_{a}$.
In Figure \ref{fig:bnumber} we see that 
the rigid bodies produce patches with higher density and pressure even close to the accretor; 
it shows that the existence of at least two cone-type shocks close to the accretor (within a distance of $5M$).
Only very close to the BH, where the streamlines are more radial, we expect that the collision
of a fluid particle with a cloud may hinder accretion on to the BH, because the obstacle
may reduce the level of (radial) collimation of the flow towards the BH.

%As we will see later, this may have important consequences for the accretion rate. 

\begin{figure}
\includegraphics[width=\columnwidth]{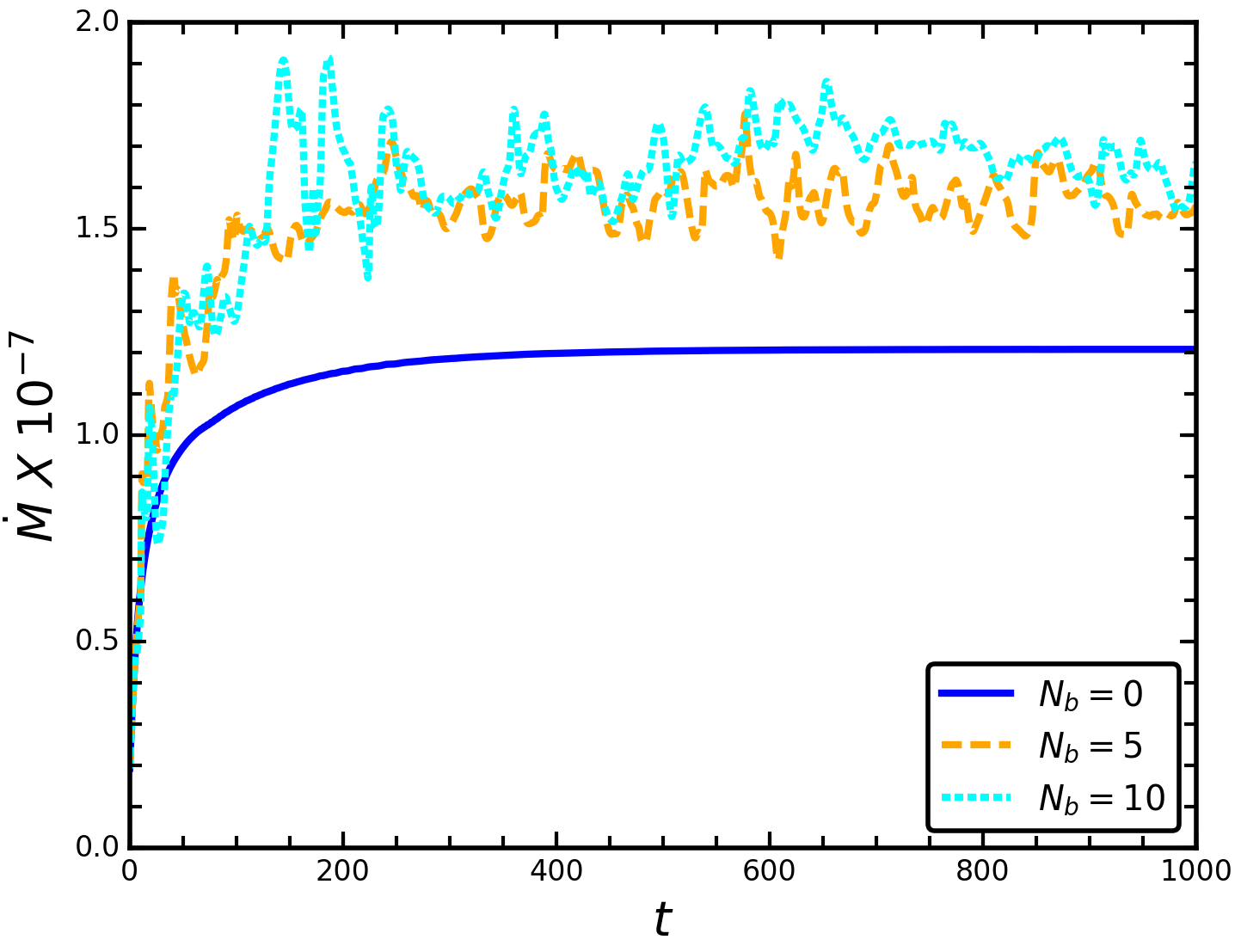} \\
\includegraphics[width=\columnwidth]{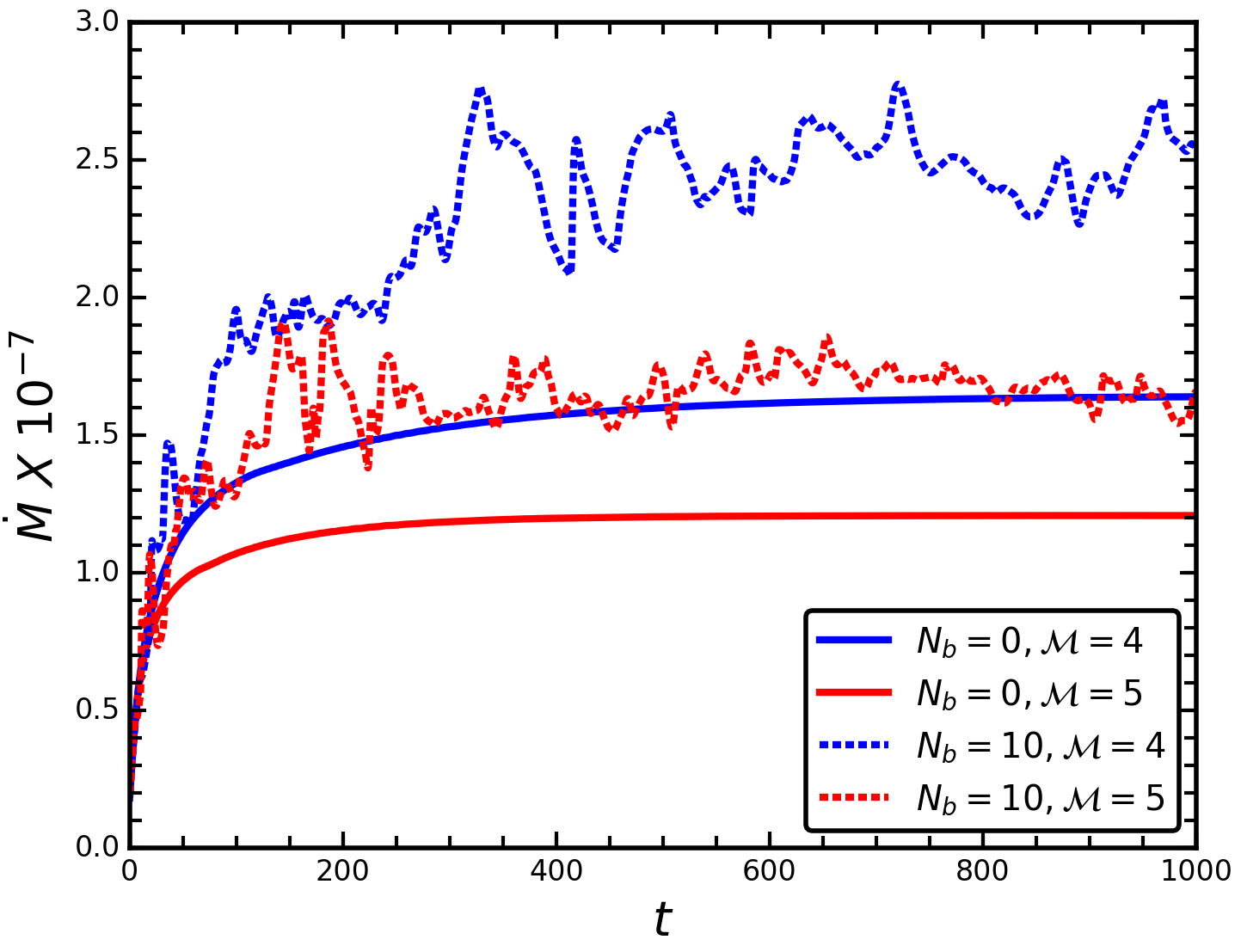} 
\caption{Mass accretion rates vs time, for different values of $N_{b}$ and ${\cal M}_{\infty}$.  
In the upper panel, we show the accretion rate for $N_{b}=0$ (solid line), $N_{b}=5$ (dashed line)
and $N_{b}=10$ (dotted line) and assuming ${\cal M}_{\infty}=5$ in all cases.
In the lower panel, we compare the mass accretion rates for ${\cal M}_{\infty}=4$ and 
${\cal M}_{\infty}=5$, with and without 
clouds.  The classical BHL accretion mass rates are pretty well behaved, but when the bubbles are included the accretion rate fluctuates around a higher value. It is worthwhile to note that the Newtonian 
formula (\ref{eq:mdotexact}) predicts $\dot{M}=0.18\times 10^{-7}$ for  ${\cal M}_{\infty}=4$, and  
$\dot{M}=0.1\times 10^{-7}$ for  ${\cal M}_{\infty}=5$. }
    \label{fig:mdotn}
\end{figure}

%%%%%
Figure \ref{fig:mdotn} shows the temporal evolution of the mass accretion rate 
for different combinations of ${\mathcal{M}}_{\infty}$ and $N_{b}$. 
In all the cases, the accretion rate starts from zero and saturates after $t\simeq 200M$.
In the classical BHL problem (i.e. $N_{b}=0$), 
the accretion rate shows a smooth behavior. On the contrary, the accretion rate fluctuates 
around a constant value in the presence of clouds. The magnitude of $\dot{M}$
is proportional to the adopted value for $\rho_{\infty}$. To be definitive, we
take $\rho_{\infty}=10^{-10}$, but the density maps can be immediately rescaled for any 
other value. For comparison, the accretion rates predicted by the Newtonian formula given
in Equation (\ref{eq:mdotexact}) are $0.18\times 10^{-7}$ for ${\cal M}_{\infty}=4$, 
and it is $0.1\times10^{-7}$ for ${\cal M}_{\infty}=5$. The accretion
rates achieved in the simulations are greater than the Newtonian values.
Table \ref{tab:params} lists the values of the relevant quantities in
physical units for some characteristic BH masses. In particular, we provide the accretion
rate found in our simulations and also the value given by the Eddington limit.

\begin{table*}
	\centering
	\caption{Summary of the mass accretion rates $\dot{M}$ found in the simulations and accretion
rates in the Eddington limit $\dot{M}_{\rm Edd}$, for different values of $M$, $\rho_{\infty}$,
$c_{s,\infty}$ and ${\cal M}_{\infty}$. 
 We give the quantities in both geometrized units in which $c=1, ~G=1,~M=1$ and 
in physical units, when using $M= 1.0 \times 10^{9}M_{\odot}, ~ 3.5 \times 10^{6}M_{\odot}$ and 
$30 M_{\odot}$, respectively. In code units, we take $\rho_{\infty}=10^{-10}$. To transform into
physical units, we rescale the gas density by a free factor $k$ given in the second row.}
\label{tab:params} 
 \begin{tabular}{|c|c|c|c|c|}\hline\hline
 Physical quantity &{Geometrized} &{Physical} & Physical & Physical   \\ 
               &{units} &  units & units & units  \\
\hline 
 $M$       &  $1.0$   &  $1\times 10^{9} ~M_{\odot}$  &  $3.5\times 10^{6} ~M_{\odot}$  &  $30.0 ~M_{\odot}$  \\
$k$   &  1.0 & $10^{-6}$ & $10^{-9}$ & $10^{-14}$  \\ 
$\rho_{\infty}$     &  $10^{-10}$ & $6.17 \times 10^{-17}$ ~g/cm$^{3}$&  $5.04 \times 10^{-15} ~g/$cm$^{3}$& $6.68 \times 10^{-10}$~g/cm$^{3}$\\ 

 $c_{s,\infty}~~$  &  $0.1$   &   $3.0 \times 10^{7}$~m/s&  $3.0 \times 10^{7}$~m/s  & $3.0 \times 10^{7}$~m/s \\ 
$\dot{M}_{{\cal M}=4,N_{b}=0}$&   $1.6\times 10^{-7}$  &$1.02 ~ M_{\odot}/$yr&  $1.02 \times 10^{-3}  ~M_{\odot}/$yr&  $1.02 \times 10^{-8}  ~M_{\odot}/$yr \\ 
 $\dot{M}_{{\cal M}=5,N_{b}=0}$  &   $1.2\times 10^{-7}$  &$0.77 ~ M_{\odot}/$yr&  $7.68 \times 10^{-4}  ~M_{\odot}/$yr&  $7.68 \times 10^{-9}  ~M_{\odot}/$yr \\ 
 $\dot{M}_{{\cal M}=4,N_{b}=10}$&   $2.5\times 10^{-7}$  &$1.60 ~ M_{\odot}/$yr&  $1.60 \times 10^{-3}  ~M_{\odot}/$yr&  $1.60 \times 10^{-8}  ~M_{\odot}/$yr \\ 
 $\dot{M}_{{\cal M}=5,N_{b}=10}$&   $1.6\times 10^{-7}$  &$1.02 ~ M_{\odot}/$yr&  $1.02 \times 10^{-3}  ~M_{\odot}/$yr&  $1.02 \times 10^{-8}  ~M_{\odot}/$yr \\ 
 $\dot{M}_{\rm Edd}$                          &    --                              & $2.20 ~M_{\odot}/$yr&  $7.71 \times 10^{-3}  ~M_{\odot}/$yr& $6.61\times 10^{-8}  ~M_{\odot}/$yr \\ 
 \hline \hline
\end{tabular}  
\end{table*}

Interestingly, for the simulations with ${\cal M}_{\infty}=5$,
just $5$ clouds are enough to enhance the accretion rate by $40\%$ 
(from $\sim 1.2\times 10^{-7}$ to $\sim 1.7 \times 10^{-7}$) over the value found in
the cloudless simulation. A similar enhancement of $50\%$ is found
for ${\cal M}_{\infty}=4$ and $N_{b}=10$. Indeed, the accretion rate over time
for $N_{b}=0$ and ${\cal M}_{\infty}=4$ is almost identical than the accretion
rate for $N_{b}=10$ and ${\cal M}_{\infty}=5$. Therefore, the inclusion of $5-10$
bodies in a flow with ${\cal M}_{\infty}=5$ increases the mass accretion rate the same 
amount than reducing ${\cal M}_{\infty}$ from $5$ to $4$.
This relative enhancement in the accretion rate is expected to increase for lower values 
of the adiabatic index $\Gamma$ because the gas becomes more compressible.

\begin{figure}
\includegraphics[width=\columnwidth]{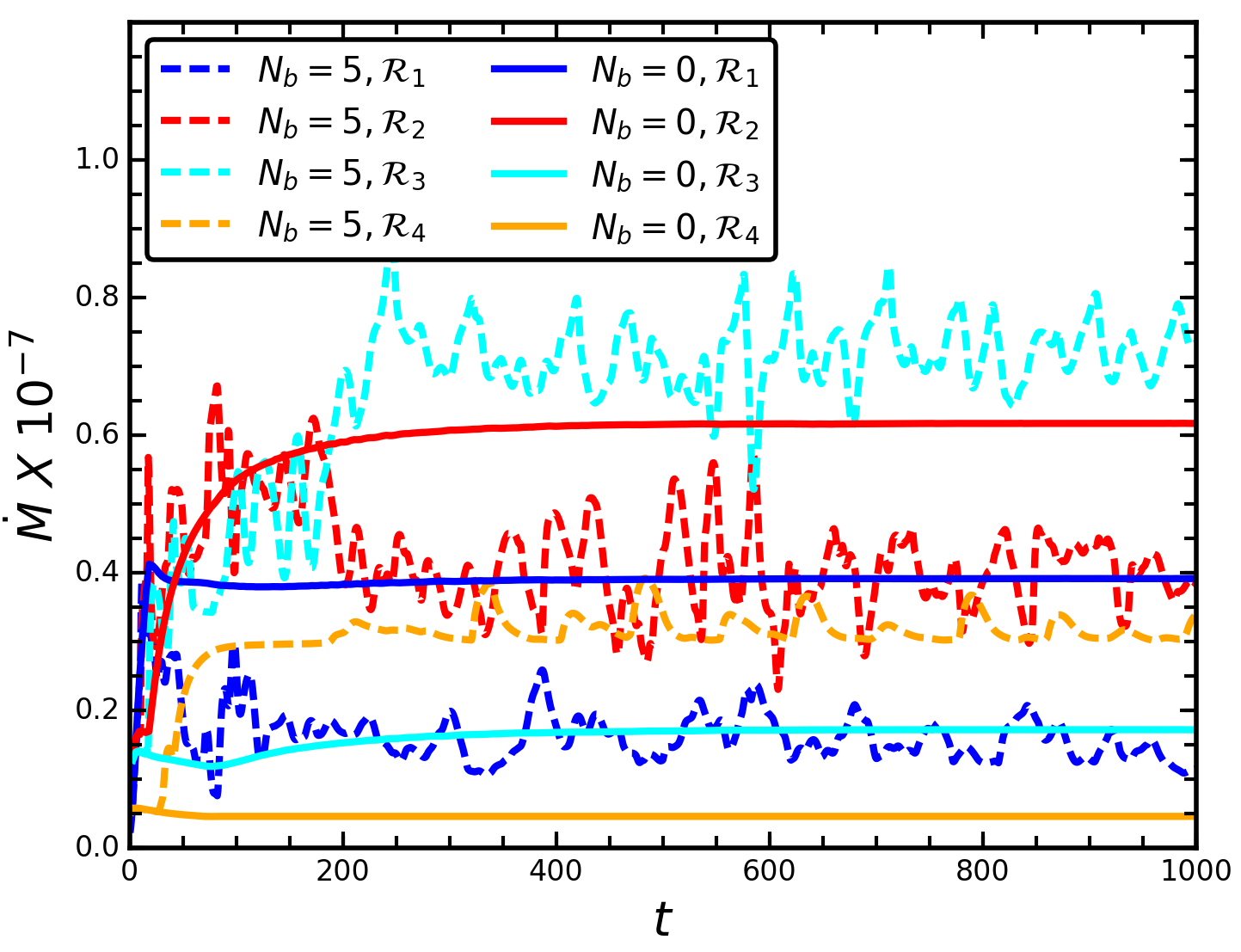}
\caption{Dependence of the mass accretion rate on the polar angle for $N_{b}=0$ (solid lines) and 
for $N_{b}=5$ (dashed lines). 
Starting from $z$ {\it -positive-axis}, which corresponds to the rear side of the accretor,
we define four regions: the region 1 (${\cal R}_{1}$) which includes 
$\theta =[0,\frac{1}{4}]\pi$, region 2 (${\cal R}_{2}$) for $\theta =[\frac{1}{4},\frac{1}{2}]\pi$, 
region 3 (${\cal R}_{3}$) for $\theta =[\frac{1}{2},\frac{3}{4}]\pi$ and region 4 (${\cal R}_{4}$)
for $\theta =[\frac{3}{4},1]\pi$. 
The accretion along the front of the BH (that is, along regions ${\cal R}_{3}$ and ${\cal R}_{4}$) is significantly
larger for $N_{b}=5$ than for $N_{b}=0$.}
    \label{fig:mreg}
\end{figure}

The mass influx as a function of the position $(r,\theta)$ contains
information about the topology of the flow. 
In the case of a naked BH (i.e. for $N_{b}=0$),
the collisionless approximation permits to find out $\rho$
and $v^{r}$ (and thereby the radial inflow of mass) in the pre-shock region %\cite[e.g.,][]{Shapiro2004}.
(Shapiro \& Teukolsky 2004). 
The mass accretion rate along the shock cone is more
uncertain and difficult to estimate analytically. 
In order to illustrate how the clouds change the structure of 
the accreting flow close to the BH,
we have computed the radial mass flux at the event horizon as a function of $\theta$, 
with and without clouds. Figure \ref{fig:mreg} shows the mass influx along four different
polar angles.  The $\theta=0$ axis corresponds to the rear axis of the accretor. It is remarkable that for the simulation with $N_{b}=0$, the accretion
rate along the polar cap at the rear of the BH (Region 1, as defined in
the caption of Figure \ref{fig:mreg}) is about $8$ times larger than the accretion rate along the 
polar cap at the front of the BH (Region 4). The reason is that the
mean density in Region 1 is about a factor $\sim 10^{2}$ larger than in Region 4,
whilst the radial inward velocity in Region 1 is only $\sim 5$ times smaller than
in Region 4.

We see that for the simulation with $N_{b}=5$, the accretion of mass along region 3 ($\pi/2\leq \theta
\leq 3\pi/4$) is $3.5$ times larger than it is when $N_{b}=0$.
On the other hand, in the simulation with $N_{b}=0$,
the accretion rate along region 1 ($0\leq \theta \leq \pi/4$)
is two times larger than it is when $N_{b}=5$. Therefore, the clumps inserted in our
simulations do alter the topology of the flow close to the BH.

\begin{figure}
\includegraphics[width=0.5\columnwidth]{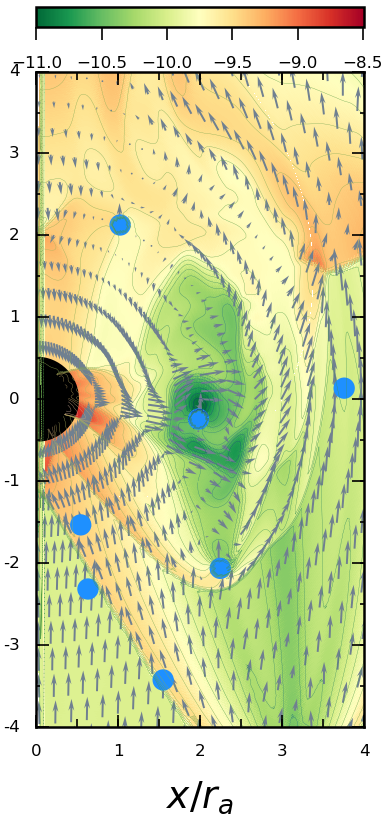} \includegraphics[width=0.5\columnwidth]{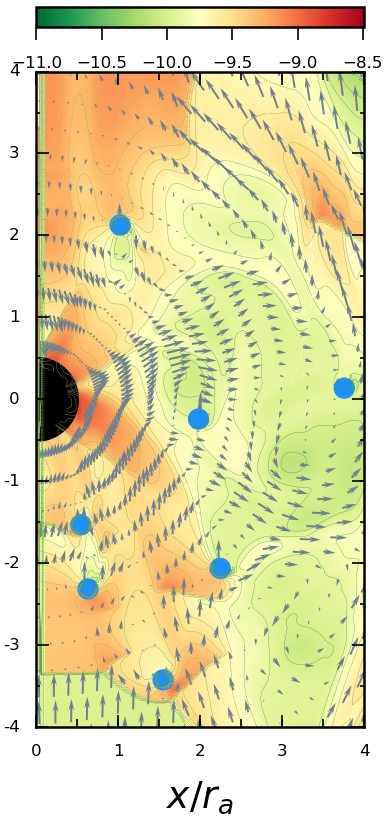}
\caption{Close up of Figure \ref{fig:bnumber} showing the morphology of the rest mass density in 
logarithmic scale near to the Schwarzschild BH
for $N_{b}=5$ (left panel) and for $N_{b}=10$ (right panel). In both cases
${\cal M}_{\infty}=5$.}
\label{fig:close}
\vspace{0.85cm}
\end{figure}

Doubling the number of rigid bodies from $N_{b}=5$ to $10$ does not change
the accretion rate significantly. One could think that the reason is that the new added
$5$ rigid bodies are located far enough from the accretor so that the flow close to the 
accretor is essentially identical. We must say that this is not the case; an enlargement of the
central region in Figure \ref{fig:close} shows that the morphology and velocity
field around the BH when $N_{b}=10$ is substantially different than they are for $N_{b}=5$. Let us consider the effect of increasing the number of clouds. The interaction of the supersonic gas with the static clouds produces shock waves.
Larger is the number density of clouds, larger is the number of shock waves
present in the gas. The shocks slow down the flow and heat the gas,
increasing the sound speed of the gas in the post-shock regions.
As shown in \citet{1983bhwd.book.....S} for spherical Bondi accretion and
in \citet{1989ApJ...336..313P} for a moving BH, a higher sound speed reduces
the accretion rate. On the other hand, a slower gas velocity respect to the BH
promotes radial accretion. What is the dominant effect? For adiabatic gas, our simulations indicate that
when increasing the number of clouds, these two effects have a compensating 
influence on the accretion rate.

\section{Summary}
\label{sec:conclusions}
The physics behind the gas accretion on to a BH is very rich and complex
because the gas may undergo various modes of accretion. These
modes depend on the (magneto-) thermodynamical properties of the gas (such
as the occurrence of thermal or magneto-thermal instabilities) but 
the accretion flow may also be affected by other energy injection sources, 
such as stellar winds or jets blown by the AGN. 
In order to isolate the different physical agents and to quantify their role, it is common
to make idealizations of the problem.
Here we have studied how BHL accretion is modified if the gas, assumed
to be adiabatic, is shocked to high temperatures due to the collision with blunt rigid objects 
located in the neighborhood of the accretor. These objects may represent cold and dense
molecular clouds, shells created by stellar winds, passing stars, or any other agent that 
results in the accreting flow becoming shocked.
We find that the morphology of the flow is sensitive to the filling factor of the clouds.
A few clouds are enough to produce a complex flow with interacting shock fronts and swirling
motions. Remarkably, the classical Mach cone at the rear of the BH is almost erased
under the presence of clouds. 
This implies that the classical picture that the accretion occurs through the $\theta=0$ axis
is not longer valid in this case because the clouds induce large
perturbations in the velocity of the flow and also in its temperature.
As a consequence, these perturbations modify the accretion rate per angle
solid in the surface of the BH.

More importantly,
we find that the presence of clouds increases the accretion rate on to the BH, even if
the medium is adiabatic. For simulations with $5$ and $10$ clouds and Mach numbers between $4$ and $5$, 
the accretion rate is about $40\%-50\%$ larger than in cloudless simulations. We attribute this 
enhancement in the accretion rate to the dragging of the flow along those streamlines that collide 
with the clouds, since they lose bulk kinetic energy and angular momentum with 
respect to the BH and this facilitates its gravitational capture because the velocity field of the
flow becomes more radial. This provides a new channel
of accretion because in the classical 
BHL scenario (in the lack of any cloud), cancellation of angular momentum
and the lost of kinetic energy can only occur along the Mach cone (i.e. the accretion column). 
We note, however, that the adiabatic condition implies that
the sound speed of the gas increases considerably in the post-shock regions and
this effect hinders accretion (e.g.  
$\dot{M}\propto c_{s,\infty}^{-3}$ in the classical Bondi accretion).
The inclusion of cooling will result in a significantly higher accretion rate.
The importance of cooling in the mass accretion rate in quasi-spherical accretion was 
highlighted by \citet{2013MNRAS.432.3401G}, who
find that the cooling enhances the accretion rate by a factor of $\sim 200$ 
as compared to the adiabatic cases. Our simulations also show some variability in the density structures and in the velocity field of the flow.  These fluctuations produce some variability in the accretion rate, 
but this is relatively small ($\sim 15\%$).

As a cautionary note, we warn that the studies 
and conclusions regarding either
the role of turbulent motions or gas cooling in Bondi accretion, i.e. when the mean
velocity of the flow relative to the BH is null \citep[e.g.][]{2011MNRAS.413.2633H,2013MNRAS.432.3401G},
cannot be generalized to the case of Bondi-Hoyle-Lyttleton accretion in a straightforward way. Nevertheless, despite the differences between accretion modes in different simulations,
the presence of turbulence, vorticity and cooling can have a substantial impact on the BH accretion rate.

In order to study how the mass accretion rate may change in a case where the gas is shocked
before accretion, we have imposed axial symmetry, implying that the
clouds are torus in 3D space. To provide a more detailed analysis is necessary
to consider 3D simulations. In addition, we have assumed as a first approximation,
the clouds to be rigid and static. If the obstacles represent dense clumps of gas,
they can suffer significant ablation by the ram pressure of the wind.
The timescale for ablation
of clumps depend on the size of the clumps, internal densities of both the wind and the clumps,
and the relative velocity. BH tidal forces may also contribute to the disruption of the clouds.
More realistic settings should include the formation, evolution and destruction of clumps in a self-consistent 
way \citep[e.g.][]{2013MNRAS.432.3401G,2014MNRAS.441.3055B,2016MNRAS.456L..20B}. In addition, in a realistic scenario, clumps should have some nonzero velocity dispersion and, thus, the relative velocity between a clump and the surrounding gas may be different from clump to clump.
What we have learned from our simulations is that the presence of obstacles around
a BH moving at Mach numbers $\sim 4$ produces significant changes in the flow
structure, so that the BHL picture of the ``accretion column'' is no longer a good
representation of the flow. We find that the mass accretion rate on to the BH may
vary by about $40-50\%$ between a quiescent regime (where there is no clumps) and another regime 
where dense substructures are present in the flow,
even if the incoming gas is adiabatic and the substructures have a small filling factor.

\section*{Acknowledgements}

The authors thank the anonymous referee for useful and insightful suggestions.
A.C.-O. gratefully acknowledges DGAPA postdoctoral grant to UNAM and European Union's Horizon 2020 Research and Innovation
Programme (Grant 671698) (call FETHPC-1-2014, project  ExaHyPE). F.D.L.-C. gratefully acknowledges financial support from Vicerrector\'ia de Investigaci\'on y Extensi\'on,  Universidad Industrial de Santander, grant number 1822 and from COLCIENCIAS, Colombia, under Grant No. 8840. 
This work has been partially supported by CONACyT project CB-165584. 
The numerical simulations were carried out on the computer TYCHO  (Posgrado de Astrof\'{\i}sica-UNAM,
Instituto de Astronom\'{\i}a-UNAM and PNPC-CONACyT).

\bibliographystyle{mnras}
%\bibliography{bibliography} 

\bsp	\label{lastpage}
\end{document}